\documentclass[twocolumn,aps,prb,epsfig,showpacs,superscriptaddress]{revtex4}

\usepackage{graphicx}

\begin{document}

\title{Enhanced magnetic moment and conductive behavior in NiFe$_2$O$_4$ spinel ultrathin films}

\author{Ulrike L\"uders}
\affiliation{Institut de Ci\`encia de Materials de Barcelona, CSIC, Campus de la UAB, 08193 Bellaterra,
Catalunya, Spain} \affiliation{LPMC-FRE2686 CNRS-ONERA, 2 avenue Edouard Belin, 31400 Toulouse, France}

\author{Manuel Bibes}
\email{manuel.bibes@ief.u-psud.fr} \affiliation{Unit\'e Mixte de Physique THALES / CNRS, Domaine de Corbeville,
91404 Orsay, France}

\affiliation{Institut d'Electronique Fondamentale, Universit\'e Paris-Sud, 91405 Orsay, France}

\author{Jean-Fran\c cois Bobo}
\affiliation{LPMC-FRE2686 CNRS-ONERA, 2 avenue Edouard Belin, 31400 Toulouse, France}

\author{Matteo Cantoni}
\author{Riccardo Bertacco}
\affiliation{INFM and L-NESS, Dipartimento di Fisica del Politecnico di Milano, via Anzani 52, 22100 Como,
Italy}

\author{Josep Fontcuberta}
\affiliation{Institut de Ci\`encia de Materials de Barcelona, CSIC, Campus de la UAB, 08193 Bellaterra,
Catalunya, Spain}

\date{\today}

\begin{abstract}

\vspace{0.5cm}

Bulk NiFe$_2$O$_4$ is an insulating ferrimagnet. Here, we report on the epitaxial growth of spinel NiFe$_2$O$_4$
ultrathin films onto SrTiO$_3$ single-crystals. We will show that - under appropriate growth conditions -
epitaxial stabilization leads to the formation of a spinel phase with magnetic and electrical properties that
radically differ from those of the bulk material : an enhanced magnetic moment (M$_S$) - about 250\% larger -
and a metallic character. A systematic study of the thickness dependence of M$_S$ allows to conclude that its
enhanced value is due to an anomalous distribution of the Fe and Ni cations among the A and B sites of the
spinel structure resulting from the off-equilibrium growth conditions and to interface effects. The relevance of
these findings for spinel- and, more generally, oxide-based heterostructures is discussed. We will argue that
this novel material could be an alternative ferromagetic-metallic electrode in magnetic tunnel junctions.

\end{abstract}

\maketitle

\section{Introduction}

Spinel ferrites \cite{brabers95} have been studied in bulk form for many years, both to understand their
magnetic behavior and correlate it to their structural properties, and to increase their performance in
high-frequency devices. Epitaxial films of spinel ferrites have not drawn such a wide attention, and have not
yet found their place in technological applications. Fe$_3$O$_4$ films have been the most studied since this
material has a high Curie temperature and is believed to be a half-metallic ferromagnet
\cite{degroot86,rudiger2002}. As such, it is of great interest in the field of spintronics
\cite{seneor99,hu2002}. Previous work on this material has allowed to identify some important differences
between the structure of the films and that of bulk samples, like the presence of antiphase boundaries (APBs)
\cite{margulies97} that alter the electrical and magnetic properties of the films \cite{eerenstein2002}.

More generally, the growth of spinel ferrite epitaxial thin films has shown that many degrees of freedom exist
to achieve modified physical properties compared to the bulk material \cite{hu2000,wakiya2004}. Recently, a
theoretical study indeed predicted new fascinating properties for NiFe$_2$O$_4$ and CoFe$_2$O$_4$, that depend
on the cationic distribution and on the electronic state of the transition metal ions of the spinel structure
\cite{szotek2004}. In some conditions, a half-metallic character is expected. This property, combined with the
high Curie temperature of these ferrites, would make them very useful as ferromagnetic electrodes in magnetic
tunnel junctions \cite{moodera95} or in other spintronics devices.

Motivated by these predictions, we have thus tried to engineer the physical properties of NiFe$_2$O$_4$ (NFO) by
growing ultrathin films of this material onto SrTiO$_3$ (STO) which has a perovskite structure. To study the
growth of spinel films onto perovskites (a group of oxides that has already shown its potential for spintronics
\cite{bowen2003,kobayashi98,bibes2003}) would pave the way towards the realization of complex architectures
integrating materials from two of the richest oxide families. Literature is sparse on this topic
\cite{venzke96,hu2002,suzuki96} and much remains to be learned concerning growth problems, and interface
properties.

In its bulk form, NFO has an inverse spinel structure and shows ferrimagnetic order below 850K. Its magnetic
structure consists of two antiferromagnetically coupled sublattices. A first sublattice is formed by
ferromagnetically ordered Fe$^{3+}$ (3d$^5$, magnetic moment (M) : 5 $\mu_B$) ions occupying the tetragonal A
sites of the spinel AB$_2$O$_4$ structure, while the second sublattice contains ferromagnetically ordered
Ni$^{2+}$ (3d$^8$, M=2 $\mu_B$) and Fe$^{3+}$ (3d$^5$, M= 5$\mu_B$) ions occupying the octahedral B sites. This
type of ordering results in a saturation magnetization of 2 $\mu_B$/f.u. (f.u. : formula unit) or 300
emu.cm$^{-3}$. Bulk NFO is an insulator \cite{austin70} and recent electronic structure calculations have
estimated gap values of 1.1 eV and 2.2 eV for spin-down and spin-up electrons, respectively \cite{szotek2004}.

In this article, we explore a range of deposition temperature in which NFO films can be grown epitaxially on
STO. Interestingly enough, we have found that nanometric films grown under reducing conditions display magnetic
and transport properties - namely an enhanced magnetic moment and a metallic behavior - that differ drastically
from the corresponding properties of bulk NFO. The magnetic moment is found to \emph{increase} upon decreasing
film thickness (t). This behavior is also in strong contrast with what is usually observed in magnetic oxides.
In manganites \cite{sun99,bibes2001e,bibes2002}, or magnetite \cite{soeya2002,eerenstein2003,bobo2002}, the
magnetization decays when the films thickness is on the order of a few unit-cells. We discuss the possible
mechanisms that may induce such an increase of the magnetization beyond bulk value and conclude that cation
inversion is the one at play here. In agreement with some recent work \cite{beach2003}, our findings illustrate
that interface effects in ultrathin Fe oxide layers may lead to some unprecedented magnetic properties. Indeed
the magnetic moment observed in our NFO films is larger than that observed in any other Fe oxide and overcomes
that reported for buried Fe-oxide layers \cite{beach2003}. In addition, we present ultraviolet photoemission
spectroscopy data that evidence the presence of a finite density-of-states at the Fermi level (E$_F$) in a NFO
thin layer, indicating a metallic character. We discuss the possible reasons for the stabilization of this new
phase of NiFe$_2$O$_4$ and give some perspectives on the possible use of this material in spintronics
experiments.

\section{Experimental details}

\subsection{Film preparation}

The NFO films were grown by off-axis target-facing-target RF sputtering in a commercial Plassys UHV chamber
\cite{bobo2002}. (001)-oriented STO single-crystals were used as substrates and heated to 800$^\circ$C before
film growth. STO has a cubic unit cell with a cell parameter a=3.905 $\rm{\AA}$. Since the cubic unit cell of
NFO has a parameter of 8.34 $\rm{\AA}$, the structural mismatch between the unit cell of NFO and a double unit
cell of STO is -6.4 \% and tends to induce a compressive strain of the film. The two NFO stoichiometric targets
were pressed pellets prepared by standard solid-state chemistry. The films were grown at temperatures ranging
from 450$^\circ$C to 550$^\circ$C in a pure Ar atmosphere at a pressure of 0.01 mbar, and with an RF power of 50
W. We have also set the deposition temperature to 550$^\circ$C and varied the deposition time, to prepare films
with nominal thickness 3, 6 and 12 nm. The growth rate was estimated from X-ray reflectometry experiments to be
0.2 nm.min$^{-1}$. This value was confirmed by high-resolution transmission electron microscopy (HRTEM)
cross-section images.

In a further step towards integration of ferrimagnetic spinels into novel functional spintronics devices, we
have also grown - at 500$^\circ$C - a nanometric (3nm) NFO film onto a LSMO/STO bilayer (LSMO :
La$_{2/3}$Sr$_{1/3}$MnO$_3$) deposited ex-situ on a STO substrate \cite{lyonnet2000}. We refer to this sample as
a LSMO/STO/NFO trilayer further on.

\subsection{Structural and magnetic properties determination}

X-ray diffraction experiments were carried out on a Philips MRD system. The thickness of the films was checked
by X-ray reflectometry. Reflection High Energy Electron Diffraction (RHEED) was operated on the substrates,
after the high temperature annealing before film growth, and at the end of deposition, along several azimuths at
an energy of 20 keV. Atomic force microscope (AFM) images were performed with a Digital Instruments system. The
magnetic properties were determined using a Quantum Design superconducting quantum interferometer device
(SQUID).

\subsection{X-ray and ultraviolet photoemission spectroscopy}

X-ray photoemission spectroscopy (XPS) and ultraviolet photoemission spectroscopy (UPS) analysis have been
performed ex-situ in a ultra high vacuum (UHV) apparatus for electron spectroscopies \cite{bertacco2004}. The
sample was the LSMO/STO/NFO trilayer. In order to reduce the surface contamination, after the introduction in the UHV system, the sample
was annealed at 500$^\circ$C for 5 minutes in presence of research graded oxygen (P=10$^{-7}$ torr). Note that this
temperature coincides with the growth temperature, so that minor changes of the sample properties are
expected. After this treatment only some traces of spurious carbon were detected by XPS on the surface. XPS and
UPS experiments have been performed employing a 150 mm hemispherical analyzer (VSW HA-150) in conjunction with a
Mg-K$_\alpha$ X-ray source and a helium I (21.2 eV) discharge lamp respectively. A Ta foil in electrical contact with
the sample has been used for calibration of the Fermi energy.

\begin{figure}[!h]
 \includegraphics[keepaspectratio=true,width=\columnwidth]{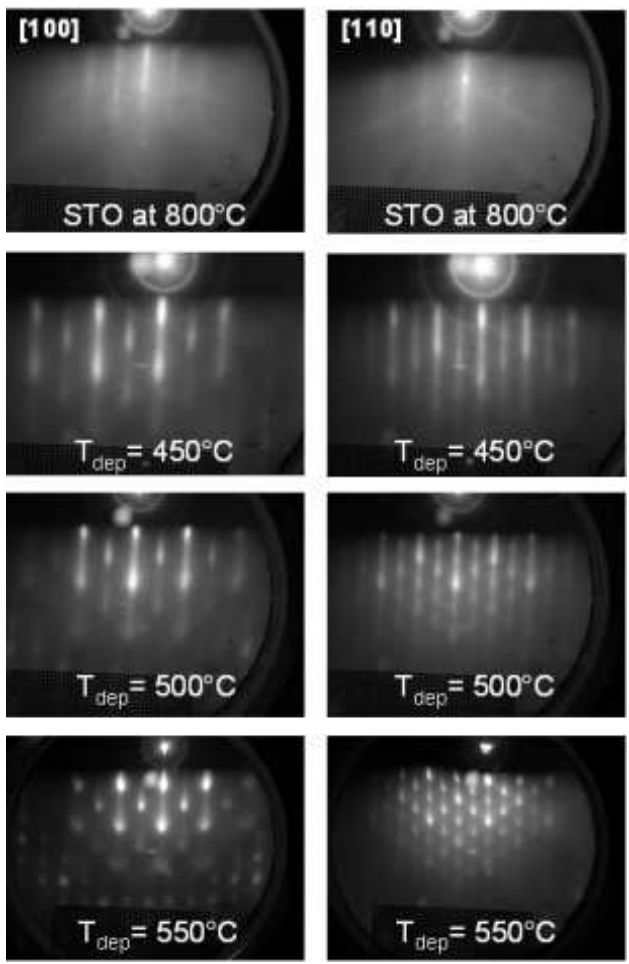}
 \caption{RHEED patterns collected on the STO substrate and on the NFO films at the end of the growth, for three
 different growth temperatures and along two azimuths corresponding to the [100] direction (left column) and
 [110] direction (right column) of STO.} \label{rheed}
\end{figure}

\vspace{0.5cm}

\section{Structural properties}

\subsection{RHEED}

RHEED patterns for 12 nm films grown at 450$^\circ$C, 500$^\circ$C and 550$^\circ$C are shown in figure
\ref{rheed}, together with those of a STO substrate prior to deposition, for two azimuths corresponding to the
[100] and [110] direction of STO. For the substrate, the observation of well-defined lattice rods and of Kikuchi
lines reflects a good surface reconstruction after the annealing process. At 450$^\circ$C, the RHEED patterns
show lattice rods along the [110] direction of STO, thus demonstrating the epitaxial quality of the film. Along
the [100] direction, they are more spotty, but the growth can be considered as almost bidimensional. The pattern
is typical of a spinel material (see for instance \cite{voogt99}), which we take as a first indication of the
formation of the NFO phase.

When T$\rm{_{dep}}$ increases, the RHEED patterns start to show periodically organized spots, evidencing an
epitaxial three-dimensional growth. The comparison of the RHEED patterns of the films with those of STO as
well as the observation of a 90$^\circ$ azimuthal period for the film RHEED patterns suggest that the films grow
cube-on-cube on the substrate. The spacing between the rods observed in the patterns for the [100] direction
decreases when increasing T$\rm{_{dep}}$, which reflects an increase of the in-plane cell parameter (a) for
films grown at higher temperatures.

Similar RHEED patterns were collected for the 3 and 6 nm-thick films which grew epitaxially in the same
three-dimensional mode as the 12 nm (deposited at 550$^\circ$C) previously discussed.

For the growth of the LSMO/STO/NFO trilayer, RHEED images were also collected. The images of the LSMO/STO
template were similar to those of the STO substrates. Those of the NFO layer grown onto the LSMO/STO template
were also identical to those of the NFO single film grown at the same temperature \cite{bibes2004}.

\begin{figure}[!h]
 \includegraphics[keepaspectratio=true,width=\columnwidth]{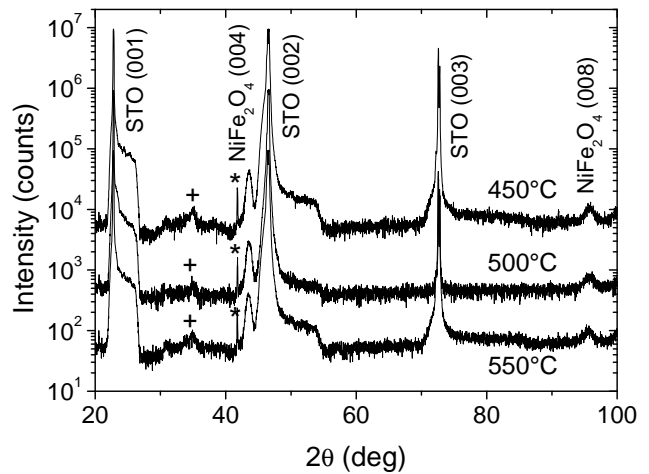}
 \caption{$\theta$-2$\theta$ scans of three 12 nm thick NFO films grown at different temperatures ;
 (*) and (+) signal reflections due to Cu K$_\beta$ and to the
plastiline holding the sample, respectively.}
 \label{2theta}
\end{figure}

\begin{figure}[!h]
 \includegraphics[keepaspectratio=true,width=\columnwidth]{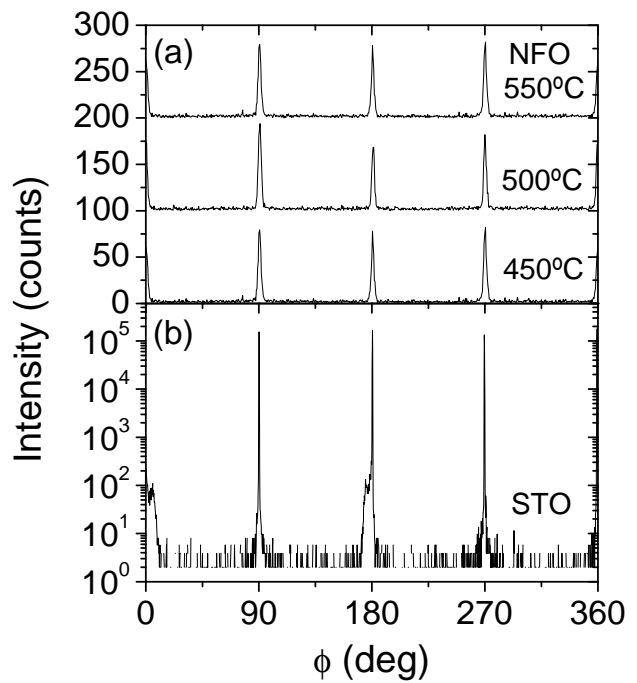}
 \caption{(a) $\phi$ scans of the (404) reflection for three 12 nm thick NFO films grown at different
 temperatures. (b) $\phi$ scans of the (202) reflection for the STO substrate.} \label{phiscans}
\end{figure}

\subsection{X-ray diffraction}

In figure \ref{2theta} we show X-ray diffraction $\theta-2\theta$ scans of these three films. Apart from the
peaks associated with the (00\emph{l}) reflections of the STO substrates, two peaks can be detected at about
43.6$^\circ$ and 95.6$^\circ$, which can be ascribed to the (004) and (008) reflections of NFO, respectively.
The absence of other reflections confirms that the films grow textured along the (001) direction. $\phi$ scans
of the (202) reflection of the STO substrate and of the (404) reflection of the films are shown in figure
\ref{phiscans}. They evidence that the films have the same in-plane texture as the substrate. From the
$\theta-2\theta$ scans and the $\phi$ scans, we conclude that the NFO films are epitaxial and grow cube-on-cube
on STO.

\begin{figure}[!h]
 \includegraphics[keepaspectratio=true,width=\columnwidth]{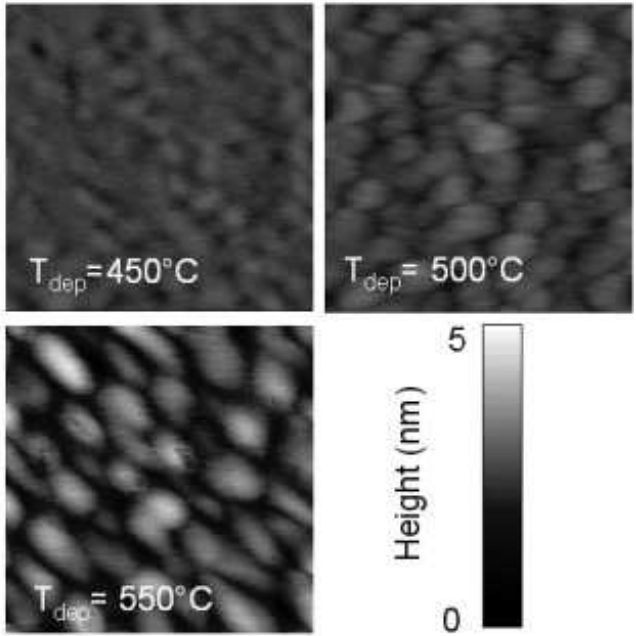}
 \caption{AFM images (200 nm $\times$ 200 nm) of 12 nm films grown at three different temperatures.}
 \label{afm}
\end{figure}

The full width at half-maximum (FWHM) of the rocking curves of the (004) reflection ($\Delta\omega_{(004)}$)
takes values in the 1.5-2$^\circ$ range and slightly decrease when T$\rm{_{dep}}$ increases. The out-of-plane
parameter (c) always takes values between 8.315 and 8.325 $\rm{\AA}$, i.e. slightly below that of the bulk. It
may reflect a smaller unit cell volume for NFO in the films, as compared to that of bulk NFO. A contraction of
the unit-cell has already been observed in NFO films grown on STO by Venzke \emph{et al} \cite{venzke96}. We
have tried to calculate the in-plane parameter (a) of the NFO films from the diffraction positions of several
asymmetric reflections. For all films, a is in the 8.3 to 8.4 $\rm{\AA}$ range, but it resulted impossible to
obtain more accurate values. Therefore, the precise strain state of the films can not be determined. Yet, since
a is not smaller than 8.3 $\rm{\AA}$, the films can be considered as virtually relaxed.

X-ray diffraction was also performed on the 3 and 6 nm films. Two small peaks could be detected and attributed
to the (004) and (008) reflection of the NFO phase in the 6 nm film, like in the 12 nm films, yielding an
out-of-plane paramater c=8.316 $\rm{\AA}$. These peaks could not be detected in the 3 nm film, likely due to the
too small amount of material. However, from RHEED (see III.A), HRTEM (see III.D) and XPS (see III.E) we are
confident that the 3 nm NFO film has a comparable structure.

\subsection{Surface morphology}

We have checked the surface morphology of the films by AFM, and the corresponding images are shown in figure
\ref{afm}. Their surface is smooth with a maximum root-mean-square (rms) roughness of 0.6 nm measured for the
film grown at 550$^\circ$C. The film grown at 450$^\circ$C is remarkably flat (rms roughness : 0.2 nm).
Grain-like features are visible, with lateral dimensions in the range of 20 nm. This morphology is in agreement
with the growth modes inferred from the RHEED patterns.

\begin{figure}[!h]
 \includegraphics[keepaspectratio=true,width=\columnwidth]{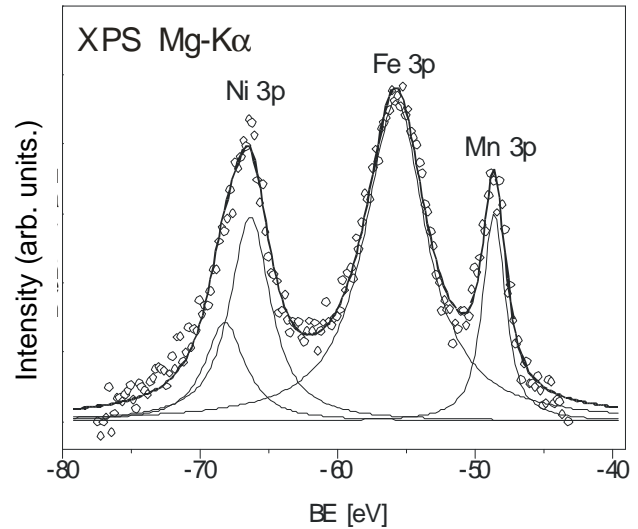}
 \caption{XPS spectrum of a 3 nm thick NFO sample grown onto a LSMO/STO(001) bilayer, in the region of Ni 3p and
 Fe 3p. The Mn 3p peak arises from the LSMO underlayer.} \label{xps}
\end{figure}

\subsection{Transmission electron microscopy and electron energy loss spectroscopy}

An extensive electron-energy loss spectroscopy (EELS) analysis was performed on a HRTEM cross section of the
LSMO/STO/NFO trilayer. These results will be published separately \cite{bibes2004}. This study revealed no
interdiffusion between the different layers and especially between STO and NFO. In this sense, the interface
between STO and NFO is abrupt and the presence of any interfacial compound can be ruled out.

\subsection{XPS}

In order to evaluate the ratio between the Ni and Fe content in the film an XPS analysis has been carried out at
10 eV of pass energy in the region of Ni 3p and Fe 3p on the LSMO/STO/NFO trilayer (see figure \ref{xps}). The
Mn 3p peak, arising from the LSMO underlying film, is also visible, so that an accurate estimate of the peak
area can be obtained only by means of an appropriate deconvolution, shown in the figure with continuous line.
For each peak we used Voigt functions with a gaussian component taking into account the experimental resolution
(FWHM of the source : 0.85 eV). For Ni 3p a doublet with the proper branching ratio (1.95) and spin-orbit
separation (1.85 eV) has been employed, while for Fe 3p and Mn 3p a single peak was used, due the smaller value
of spin-orbit interaction for these levels \cite{scofield76}. From this analysis we obtain a ratio of
0.50$\pm$0.05 between the Ni 3p and Fe 3p areas normalized to the photoemission cross sections. As the kinetic
energy of photoelectrons corresponding to Ni 3p and Fe 3p peaks is essentially the same (1182.9 eV and 1193.5
eV) the influence of the electron escape depth and analyzer transmission on the peak intensity is also the same,
so that the ratio of the areas normalized to the cross sections directly reflects the ratio between the Ni and
Fe concentrations. We can then conclude that the experimental ratio between the Ni and Fe content in the
investigated NFO film is in good agreement with the expected stoichiometry: NiFe$_2$O$_4$.

\section{Magnetic properties}

\subsection{Influence of the growth temperature}

The magnetization of the films was measured at 10K after zero-field cooling, with the magnetic field applied in
the plane, along the [100] direction. In addition to the ferromagnetic signal of the NFO film, a large negative
slope was always observed at high-field and ascribed to the diamagnetism of the STO substrate. However, we can
not exclude that this high-field part also contains some contribution from the film, due for instance to
spin-disorder (giving rise to a positive dM/dH slope). If present, this contribution could not be discriminated
from that arising from the substrate. To retrieve the ferromagnetic signal coming from the film, we have fitted
the linear field dependence of the magnetization in the 30 to 50 kOe range, and subtracted this contribution
from the experimental data. This contribution was almost the same for all the films we have measured and roughly
equal to that given by a virgin STO substrate.

\begin{figure}[!h]
 \includegraphics[keepaspectratio=true,width=\columnwidth]{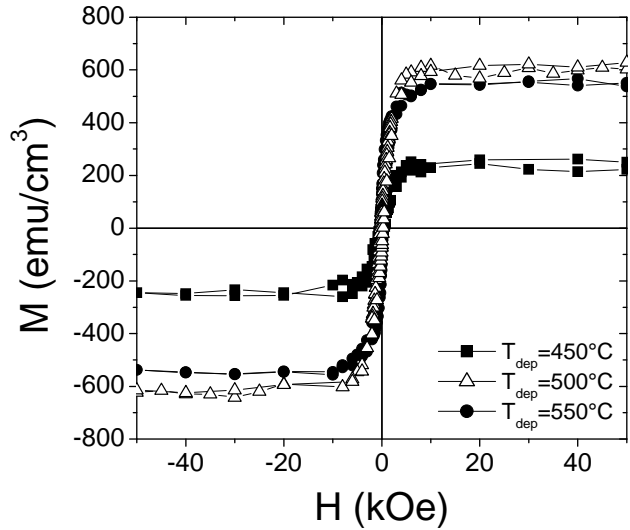}
 \caption{Hysteresis cycles measured at 10K and corrected from the high-field contribution of the substrate, for
 three 12 nm films grown at different temperatures.} \label{m_h_tdep}
\end{figure}

The corrected hysteresis cycles for the three films grown at 450$^\circ$C, 500$^\circ$C and 550$^\circ$C are
presented in figure \ref{m_h_tdep}. The film grown at 450$^\circ$C shows a saturation magnetization M$_S$ of
about 250 emu.cm$^{-3}$, slightly smaller than that of the bulk (300 emu.cm$^{-3}$). The two other films have
M$_S$ values in the 600 emu.cm$^{-3}$ range, far above the bulk moment. This is a priori surprising since no
parasite ferromagnetic phases could be detected by the different characterization techniques employed. However,
the observation of an excessively high magnetic moment has already been reported in homoepitaxial NFO films
\cite{venzke96}.

\subsection{Thickness dependence}

\begin{figure}[!h]
 \includegraphics[keepaspectratio=true,width=\columnwidth]{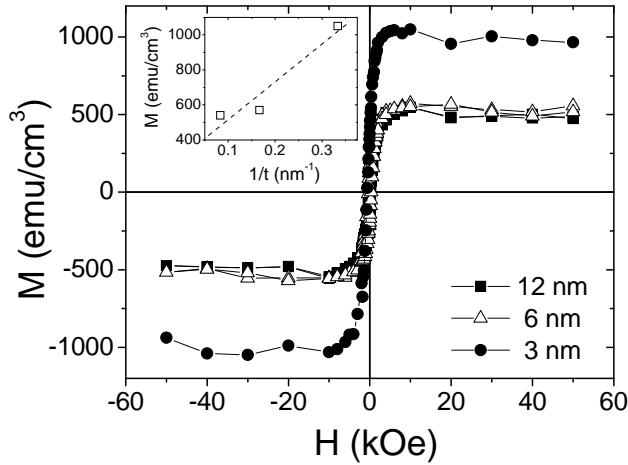}
 \caption{Hysteresis cycles measured at 10K and corrected from the high-field contribution of the substrate for
 three films with different thickness grown at 550$^\circ$C. Inset : Variation of the saturation magnetization
 with the inverse of the film thickness. The dotted line is a linear fit.} \label{m_h_thick}
\end{figure}

In order to get information on the interface properties of the high-magnetization NFO films grown on STO, we
have measured the magnetization of the 3, 6 and 12 nm films grown at 550$^\circ$C. Hysteresis cycles measured at
10K after zero-field cooling are shown in figure \ref{m_h_thick} (after correction). All the films show a
saturation magnetization larger than that of bulk NFO, and this effect is especially pronounced for the 3 nm
film, which has a moment of about 1050 emu.cm$^{-3}$. This latter value has be confirmed within an error of 15\%
for several identical films. The results are summarized in the inset of figure \ref{m_h_thick} in which we plot
the saturation magnetization \emph{vs} the inverse of film thickness. A roughly linear dependence is obtained,
suggesting that the increase of M is promoted at the NFO/STO interface.

\section{Valence-band spectroscopy}

\begin{figure}[!h]
\includegraphics[keepaspectratio=true,width=\columnwidth]{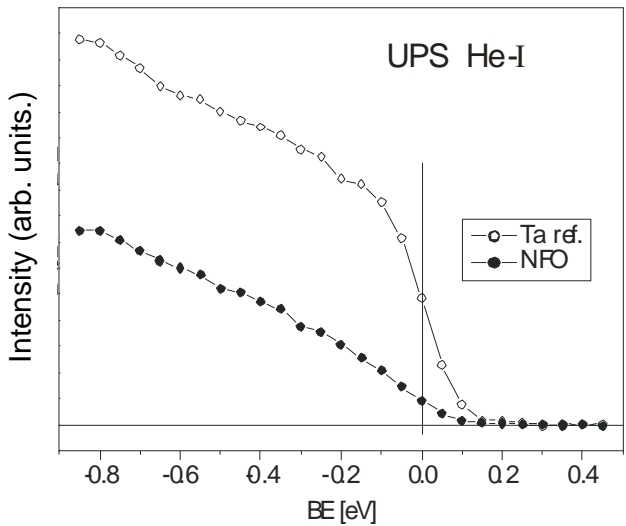}
\caption{UPS spectra from the NFO sample already used for XPS investigation and from a Ta foil in electrical
 contact with the sample.}
\label{ups}
\end{figure}

In order to explore the electronic structure of these nanometric NFO films grown in Ar atmosphere, a UPS
analysis of the valence band has been carried out on the same trilayer employed for EELS and XPS. The
corresponding UPS spectrum is shown in figure \ref{ups} together with a reference spectrum from a Ta foil in
electrical contact with NFO. No surface long-range order was detected on this sample (via low-energy electron
diffraction analysis), so that the UPS spectra essentially reflect the sample density of states, instead of
vertical transitions between states of the full band structure. As the onset of the UPS spectrum from NFO is
placed exactly at the Fermi level, as determined from the edge of the Ta spectrum, the sample behaves as a
conductor. Within the finite resolution of our set up ($\sim$100 meV), in fact, a small but finite density of
states is present at the Fermi level thus ensuring electrical conduction.

\section{Discussion}

As previously mentioned, a magnetic moment 25 \% greater than that of the bulk has already been reported for NFO
films by Venzke \emph{et al} \cite{venzke96}. However, the effect that we describe here is much more spectacular
since is corresponds to an enhancement of up to 250 \% and is clearly promoted by the reduction of thickness.
Several explanations can be invoked to account for our observation.

First, this large magnetic moment could be due to the presence of parasite phases. Possible candidates include
Fe or Ni oxides and metallic alloys of Ni and Fe. However, no Fe or Ni oxide has a magnetic moment as large as
the one we measure for the thinner NFO film. Furthermore, X-ray diffraction, RHEED, HRTEM and EELS clearly
do not support the presence of parasite phases and in fact provide evidence that the largest fraction of the
sample volume, if not all, corresponds to a spinel phase.

Second, one must consider Fe vacancies on A sites as a factor resulting in an enhanced magnetic moment.
Nevertheless, this option can be ruled out as the Ni/Fe ratio we measure by XPS is 0.50$\pm$0.05, that is the
amount of Fe vacancies, if non-zero, is very small.

A third possibility could be the presence of oxygen vacancies in our films. In that case, the valence of some Fe
ions would likely be 2+ instead of 3+ (Ni$^+$ is extremely improbable). In the picture of a bulk-like cationic
distribution on the A and B sites, and assuming that all the Fe ions lying at A sites are 2+, with a moment of 4
$\mu_B$, the total magnetic moment rises to 3$\mu_B$/f.u., i.e. 450 emu.cm$^{-3}$. This is clearly not enough to
explain our data.

A fourth possibility, already invoked by Venzke \emph{et al} \cite{venzke96}, is a change in the cation
distribution. Indeed, if all the Ni$^{2+}$ replace the Fe$^{3+}$ at A sites, and vice-versa, ending at a A
sublattice fully filled with Ni$^{2+}$ and a B sublattice fully consisting of Fe$^{3+}$ (normal spinel
structure), the total magnetic moment can increase up to 8 $\mu_B$/f.u. (1200 emu.cm$^{-3}$). This scenario can
explain our data and especially the very large magnetic moment of 1050 emu.cm$^{-3}$ measured for the 3 nm film.
This value would correspond to 83\% of inversion. Such a high degree of inversion could indicate that close to
interface, the normal form is stabilized over few nm. We can not exclude however, that deeper into the film the
inverse structure of the bulk NiFe$_2$O$_4$ is already formed. This could be an appealing possibility, although
the driving force for such stabilization cannot be related to strain effects as all films (thicker that 3 nm)
appear to be relaxed.

As observed by UPS, the films have a metallic character. We might expect that conduction occurs between
mixed-valent Fe ions located in the B sublattice, as happens in magnetite. In this scenario, some of the Fe ions
must be 2+, which can be expected as the films are grown in pure Ar atmosphere. The presence of some Fe$^{2+}$
ions in the B sublattice would induce some decrease of the magnetic moment from the value of 8 $\mu_B$/f.u.
expected for stoichiometric normal NFO.

In the following, we discuss the possible origin of the anomalous cationic distribution in our films. In spinel
ferrites, the distribution of the ions on the A and B sites is determined by the total energy of the crystal,
which depends on a number of factors like the ion size, the Coulomb energy of the ions in the lattice, etc. The
difference in energy $\Delta$ between the normal and inverse spinel structure \cite{cormack88} ranges from some hundreds
of mev to more that 1 eV. In the case of a small $\Delta$, a mixed inverse/normal structure can
occur.

In NiFe$_2$O$_4$, the inverse spinel structure is more stable than the normal structure by $\Delta$=1.6 eV
\cite{cormack88}. Therefore, deviations from bulk-like cation distribution are less likely to occur than in
other ferrites \cite{tsukimura97}. However, substantial levels (up to $\sim$ 10\%) of cation inversion have been
reported in NFO quenched crystals \cite{robertson66}. This shows that in off-equilibrium conditions, cations can
be stabilized in energetically unfavorable sites, ending at a mixed normal-inverse spinel structure. Our films
have been grown by RF sputtering and the ions in an RF plasma are usually highly energetic, as compared to what
happens in the case of molecular beam epitaxy for instance. Thus, it is very likely that the material deposited
onto the substrate is formed in conditions far from the thermodynamic equilibrium. As a result, one might expect
Fe and Ni ions to be randomly distributed among the A and B sites. This would yield a magnetic moment of 4
$\mu_B$/f.u. or 600 emu.cm$^{-3}$, which is very close to the value obtained for the 12 nm films grown at 500
and 550$^\circ$C.

Besides, several studies on nanoparticles \cite{kim2002,chinnasamy2001,zhou2002} have demonstrated that cationic
inversion is promoted at surfaces. This might indicate that the value of $\Delta$ strongly decreases at a
surface and may even change sign. In our films, the magnetic moment increases as thickness decreases, indicating
some interface effect. We therefore argue that the similar effect observed at nanoparticles surfaces is
occurring at the NFO/STO interface, resulting in a large cationic disorder and an enhanced magnetic moment.

Finally we would like to drive the attention to the fact that the understanding of magnetism of the metallic and
ferromagnetic nanometric NFO films we have been able to stabilize, will require a radically different approach
than usually used to describe ferromagnetism in spinels. Notice that in bulk NFO the insulating character make
superexchange interaction model appropriate to describe ferrimagnetic ordering. However, in the present case of
metallic NFO, different approaches, such as itinerant Ruderman-Kittel-Kasuya-Yosida (RKKY) exchange or even
double-exchange ferromagnetic coupling mechanisms should be invoked to describe the nature of ferromagnetic
ordering.

\section{Conclusion}

In summary, we have grown NiFe$_2$O$_4$ epitaxial thin films on (001)-oriented STO substrates in pure argon
atmosphere, with thicknesses ranging from 3 to 12 nm. The films grow cube-on-cube with no evidence for parasite
phases or interdiffusion, and a relatively smooth surface state (roughness $\sim$ 0.5 nm). Surprisingly, their
magnetic moment is larger than that of the bulk compound, and increases up to 1050 emu.cm$^{-3}$ (i.e. almost 4
times the bulk value) as thickness decreases. As we rule out the presence of parasite phases and since the Ni/Fe
ratio as measured by XPS is 0.50$\pm$0.05, this enhanced magnetic moment can only be explained by a cation
distribution different from that of the bulk, the thinnest film (3 nm) having 83\% degree of cation inversion.
We argue that this anomalous distribution arises from the growth process and from interface effects, in a
similar way as what occurs at the surface of NFO nanoparticles. In analogy to recent results on nanometric
buried Fe oxide layers\cite{beach2003}, our findings thus show that unexpectedly large magnetic moments can be
obtained through interface effects in Fe oxides.

In addition, we show by UPS that the NFO films have a finite density of states at the Fermi level, which
indicates a conductive behavior, in contrast with the insulating properties of bulk NFO. We have thus managed to
prepare a new phase of NiFe$_2$O$_4$, which is not stable in bulk form, by means of epitaxial growth on a
mismatched substrate, in a pure argon atmosphere. The microscopic description of ferromagnetic coupling in these
new oxides is still open, but the role of itinerant carriers associated to the observed metallic behavior cannot
be neglected. Indeed, a carrier-mediated ferromagnetic coupling could be active. The possibility of a high
spin-polarization in this new metallic ferromagnet is also exciting and would deserve further experiments. This
material could thus be useful, for instance, as a ferromagnetic electrode in magnetic tunnel junctions. More
generally, our study provides a way to design new magnetic materials which might be of great interest for
spintronics and other fields. Spinel ferrites epitaxial thin films may appear as a new playground for materials
scientists.

\vspace{0.5cm}

This work was partially supported by the France-Spain PICASSO program, the CICYT of the Spanish Government
(project MAT2002-04551-cO3) and FEDER. We wish to thank Jean-Luc Maurice for HRTEM analysis as well as Pierre
Baul\`es, Thomas Blon and Christophe Gatel for their assistance in sample characterization. We are also grateful
to Agn\`es Barth\'el\'emy, Karim Bouzehouane, Niels Keller, Philippe Tailhades and Dzidka Szotek for helpful
discussions.

\bibliographystyle{prsty}

\begin{thebibliography}{35}
\expandafter\ifx\csname natexlab\endcsname\relax\def\natexlab#1{#1}\fi \expandafter\ifx\csname
bibnamefont\endcsname\relax
  \def\bibnamefont#1{#1}\fi
\expandafter\ifx\csname bibfnamefont\endcsname\relax
  \def\bibfnamefont#1{#1}\fi
\expandafter\ifx\csname citenamefont\endcsname\relax
  \def\citenamefont#1{#1}\fi
\expandafter\ifx\csname url\endcsname\relax
  \def\url#1{\texttt{#1}}\fi
\expandafter\ifx\csname urlprefix\endcsname\relax\def\urlprefix{URL }\fi \providecommand{\bibinfo}[2]{#2}
\providecommand{\eprint}[2][]{\url{#2}}

\bibitem[{\citenamefont{{V.A.M. Brabers}}(1995)}]{brabers95}
\bibinfo{author}{\bibnamefont{{V.A.M. Brabers}}},
  \emph{\bibinfo{title}{Handbook of Magnetic Materials}},
  vol.~\bibinfo{volume}{8} (\bibinfo{year}{1995}).

\bibitem[{\citenamefont{{R.A. de Groot} and {K.H.J.
  Buschow}}(1986)}]{degroot86}
\bibinfo{author}{\bibnamefont{{R.A. de Groot}}} \bibnamefont{and}
  \bibinfo{author}{\bibnamefont{{K.H.J. Buschow}}}, \bibinfo{journal}{J. Magn.
  Magn. Mater.} \textbf{\bibinfo{volume}{54-57}}, \bibinfo{pages}{1377}
  (\bibinfo{year}{1986}).

\bibitem[{\citenamefont{{Yu.S. Dedkov} et~al.}(2002)\citenamefont{{Yu.S.
  Dedkov}, R\"udiger, and G\"untherodt}}]{rudiger2002}
\bibinfo{author}{\bibnamefont{{Yu.S. Dedkov}}},
  \bibinfo{author}{\bibfnamefont{U.}~\bibnamefont{R\"udiger}},
  \bibnamefont{and}
  \bibinfo{author}{\bibfnamefont{G.}~\bibnamefont{G\"untherodt}},
  \bibinfo{journal}{Phys. Rev. B} \textbf{\bibinfo{volume}{65}},
  \bibinfo{pages}{064417} (\bibinfo{year}{2002}).

\bibitem[{\citenamefont{Seneor et~al.}(1999)\citenamefont{Seneor, Fert,
  Maurice, Montaigne, Petroff, and Vaur\`es}}]{seneor99}
\bibinfo{author}{\bibfnamefont{P.}~\bibnamefont{Seneor}},
  \bibinfo{author}{\bibfnamefont{A.}~\bibnamefont{Fert}},
  \bibinfo{author}{\bibfnamefont{J.-L.} \bibnamefont{Maurice}},
  \bibinfo{author}{\bibfnamefont{F.}~\bibnamefont{Montaigne}},
  \bibinfo{author}{\bibfnamefont{F.}~\bibnamefont{Petroff}}, \bibnamefont{and}
  \bibinfo{author}{\bibfnamefont{A.}~\bibnamefont{Vaur\`es}},
  \bibinfo{journal}{Appl. Phys. Lett.} \textbf{\bibinfo{volume}{74}},
  \bibinfo{pages}{4017} (\bibinfo{year}{1999}).

\bibitem[{\citenamefont{Hu and Suzuki}(2002)}]{hu2002}
\bibinfo{author}{\bibfnamefont{G.}~\bibnamefont{Hu}} \bibnamefont{and}
  \bibinfo{author}{\bibfnamefont{Y.}~\bibnamefont{Suzuki}},
  \bibinfo{journal}{Phys. Rev. Lett.} \textbf{\bibinfo{volume}{89}},
  \bibinfo{pages}{276601} (\bibinfo{year}{2002}).

\bibitem[{\citenamefont{{D.T. Margulies} et~al.}(1997)\citenamefont{{D.T.
  Margulies}, {F.T. Parker}, {M.L. Rudee}, {F.E. Spada}, {J.N. Chapman}, {P.R.
  Aitchison}, and {A.E. Berkowitz}}}]{margulies97}
\bibinfo{author}{\bibnamefont{{D.T. Margulies}}},
  \bibinfo{author}{\bibnamefont{{F.T. Parker}}},
  \bibinfo{author}{\bibnamefont{{M.L. Rudee}}},
  \bibinfo{author}{\bibnamefont{{F.E. Spada}}},
  \bibinfo{author}{\bibnamefont{{J.N. Chapman}}},
  \bibinfo{author}{\bibnamefont{{P.R. Aitchison}}}, \bibnamefont{and}
  \bibinfo{author}{\bibnamefont{{A.E. Berkowitz}}}, \bibinfo{journal}{Phys.
  Rev. Lett.} \textbf{\bibinfo{volume}{79}}, \bibinfo{pages}{5162}
  (\bibinfo{year}{1997}).

\bibitem[{\citenamefont{Eerenstein et~al.}(2002)\citenamefont{Eerenstein,
  {T.T.M. Palstra}, Hibma, and Celotto}}]{eerenstein2002}
\bibinfo{author}{\bibfnamefont{W.}~\bibnamefont{Eerenstein}},
  \bibinfo{author}{\bibnamefont{{T.T.M. Palstra}}},
  \bibinfo{author}{\bibfnamefont{T.}~\bibnamefont{Hibma}}, \bibnamefont{and}
  \bibinfo{author}{\bibfnamefont{S.}~\bibnamefont{Celotto}},
  \bibinfo{journal}{Phys. Rev. B} \textbf{\bibinfo{volume}{66}},
  \bibinfo{pages}{201101(R)} (\bibinfo{year}{2002}).

\bibitem[{\citenamefont{Hu et~al.}(2000)\citenamefont{Hu, Choi, Eom, Harris,
  and Suzuki}}]{hu2000}
\bibinfo{author}{\bibfnamefont{G.}~\bibnamefont{Hu}},
  \bibinfo{author}{\bibfnamefont{J.H.}~\bibnamefont{Choi}},
  \bibinfo{author}{\bibfnamefont{C.B.}~\bibnamefont{Eom}},
  \bibinfo{author}{\bibfnamefont{V.G.}~\bibnamefont{Harris}}, \bibnamefont{and}
  \bibinfo{author}{\bibfnamefont{Y.}~\bibnamefont{Suzuki}},
  \bibinfo{journal}{Phys. Rev. B} \textbf{\bibinfo{volume}{62}},
  \bibinfo{pages}{R779} (\bibinfo{year}{2000}).

\bibitem[{\citenamefont{Wakiya et~al.}(2002)\citenamefont{Wakiya, Shinozaki, and Mizutani}}]{wakiya2004}
\bibinfo{author}{\bibfnamefont{N.}~\bibnamefont{Wakiya}},
  \bibinfo{author}{\bibfnamefont{K.}~\bibnamefont{Shinozaki}}, \bibnamefont{and}
  \bibinfo{author}{\bibfnamefont{N.}~\bibnamefont{Mizutani}},
  \bibinfo{journal}{Appl. Phys. Lett.} \textbf{\bibinfo{volume}{85}},
  \bibinfo{pages}{1199} (\bibinfo{year}{2004}).

\bibitem[{\citenamefont{Szotek}(2004)}]{szotek2004}
\bibinfo{author}{\bibfnamefont{Z.}~\bibnamefont{Szotek}},
  \bibinfo{journal}{Presented at the International Conference on
  Nanospintronics, Kyoto, May 24-28}  (\bibinfo{year}{2004}).

\bibitem[{\citenamefont{Moodera et~al.}(1995)\citenamefont{Moodera, Kinder,
  Wong, and Meservey}}]{moodera95}
\bibinfo{author}{\bibfnamefont{J.S.}~\bibnamefont{Moodera}},
  \bibinfo{author}{\bibfnamefont{L.R.}~\bibnamefont{Kinder}},
  \bibinfo{author}{\bibfnamefont{T.M.}~\bibnamefont{Wong}}, \bibnamefont{and}
  \bibinfo{author}{\bibfnamefont{R.}~\bibnamefont{Meservey}},
  \bibinfo{journal}{Phys. Rev. Lett.} \textbf{\bibinfo{volume}{74}},
  \bibinfo{pages}{3273} (\bibinfo{year}{1995}).

\bibitem[{\citenamefont{Bowen et~al.}(2003)\citenamefont{Bowen, Bibes,
  Barth\'el\'emy, Contour, Anane, Lema\^{\i}tre, and Fert}}]{bowen2003}
\bibinfo{author}{\bibfnamefont{M.}~\bibnamefont{Bowen}},
  \bibinfo{author}{\bibfnamefont{M.}~\bibnamefont{Bibes}},
  \bibinfo{author}{\bibfnamefont{A.}~\bibnamefont{Barth\'el\'emy}},
  \bibinfo{author}{\bibfnamefont{J.-P.} \bibnamefont{Contour}},
  \bibinfo{author}{\bibfnamefont{A.}~\bibnamefont{Anane}},
  \bibinfo{author}{\bibfnamefont{Y.}~\bibnamefont{Lema\^{\i}tre}},
  \bibnamefont{and} \bibinfo{author}{\bibfnamefont{A.}~\bibnamefont{Fert}},
  \bibinfo{journal}{Appl. Phys. Lett.} \textbf{\bibinfo{volume}{82}},
  \bibinfo{pages}{233} (\bibinfo{year}{2003}).

\bibitem[{\citenamefont{Kobayashi et~al.}(1998)\citenamefont{Kobayashi, Kimura,
  Sawada, Terakura, and Tokura}}]{kobayashi98}
\bibinfo{author}{\bibfnamefont{K.}~\bibnamefont{Kobayashi}},
  \bibinfo{author}{\bibfnamefont{T.}~\bibnamefont{Kimura}},
  \bibinfo{author}{\bibfnamefont{H.}~\bibnamefont{Sawada}},
  \bibinfo{author}{\bibfnamefont{K.}~\bibnamefont{Terakura}}, \bibnamefont{and}
  \bibinfo{author}{\bibfnamefont{Y.}~\bibnamefont{Tokura}},
  \bibinfo{journal}{Nature} \textbf{\bibinfo{volume}{395}},
  \bibinfo{pages}{677} (\bibinfo{year}{1998}).

\bibitem[{\citenamefont{Bibes et~al.}(2003)\citenamefont{Bibes, Bouzehouane,
  Besse, Barth\'el\'emy, S.Fusil, Bowen, Seneor, Contour, and
  Fert}}]{bibes2003}
\bibinfo{author}{\bibfnamefont{M.}~\bibnamefont{Bibes}},
  \bibinfo{author}{\bibfnamefont{K.}~\bibnamefont{Bouzehouane}},
  \bibinfo{author}{\bibfnamefont{M.}~\bibnamefont{Besse}},
  \bibinfo{author}{\bibfnamefont{A.}~\bibnamefont{Barth\'el\'emy}},
  \bibinfo{author}{\bibnamefont{S.Fusil}},
  \bibinfo{author}{\bibfnamefont{M.}~\bibnamefont{Bowen}},
  \bibinfo{author}{\bibfnamefont{P.}~\bibnamefont{Seneor}},
  \bibinfo{author}{\bibfnamefont{J.-P.} \bibnamefont{Contour}},
  \bibnamefont{and} \bibinfo{author}{\bibfnamefont{A.}~\bibnamefont{Fert}},
  \bibinfo{journal}{Appl. Phys. Lett.} \textbf{\bibinfo{volume}{83}},
  \bibinfo{pages}{2629} (\bibinfo{year}{2003}).

\bibitem[{\citenamefont{Venzke et~al.}(1996)\citenamefont{Venzke, van Dover,
  Philips, Gyorgy, Siegrist, Chen, Werder, Fleming, Felder, Coleman
  et~al.}}]{venzke96}
\bibinfo{author}{\bibfnamefont{S.}~\bibnamefont{Venzke}},
  \bibinfo{author}{\bibfnamefont{R.B.}~\bibnamefont{van Dover}},
  \bibinfo{author}{\bibfnamefont{J.M.}~\bibnamefont{Philips}},
  \bibinfo{author}{\bibfnamefont{E.M.}~\bibnamefont{Gyorgy}},
  \bibinfo{author}{\bibfnamefont{T.}~\bibnamefont{Siegrist}},
  \bibinfo{author}{\bibfnamefont{C.-H.} \bibnamefont{Chen}},
  \bibinfo{author}{\bibfnamefont{D.}~\bibnamefont{Werder}},
  \bibinfo{author}{\bibfnamefont{R.M.}~\bibnamefont{Fleming}},
  \bibinfo{author}{\bibfnamefont{R.J.}~\bibnamefont{Felder}},
  \bibinfo{author}{\bibfnamefont{E.}~\bibnamefont{Coleman}},
  \bibnamefont{and} \bibinfo{author}{\bibfnamefont{R.}~\bibnamefont{Opila}},
  \bibinfo{journal}{J. Mater. Res} \textbf{\bibinfo{volume}{11}},
  \bibinfo{pages}{1187} (\bibinfo{year}{1996}).

\bibitem[{\citenamefont{Suzuki et~al.}(1996)\citenamefont{Suzuki, {R.B. van
  Dover}, {E.M. Gyorgy}, {J.M. Philips}, Korenivksi, {D.J. Wender}, {C.H.
  Chen}, {R.J. Cava}, {J.J. Krajewski}, {W.F. Peck Jr.} et~al.}}]{suzuki96}
\bibinfo{author}{\bibfnamefont{Y.}~\bibnamefont{Suzuki}},
  \bibinfo{author}{\bibnamefont{{R.B. van Dover}}},
  \bibinfo{author}{\bibnamefont{{E.M. Gyorgy}}},
  \bibinfo{author}{\bibnamefont{{J.M. Philips}}},
  \bibinfo{author}{\bibfnamefont{V.}~\bibnamefont{Korenivksi}},
  \bibinfo{author}{\bibnamefont{{D.J. Wender}}},
  \bibinfo{author}{\bibnamefont{{C.H. Chen}}},
  \bibinfo{author}{\bibnamefont{{R.J. Cava}}},
  \bibinfo{author}{\bibnamefont{{J.J. Krajewski}}},
  \bibinfo{author}{\bibnamefont{{W.F. Peck Jr.}}},
  \bibnamefont{and} \bibinfo{author}{\bibfnamefont{K.B.}~\bibnamefont{Do}},
  \bibinfo{journal}{Appl. Phys. Lett.} \textbf{\bibinfo{volume}{68}},
  \bibinfo{pages}{714} (\bibinfo{year}{1996}).

\bibitem[{\citenamefont{{I.G. Austin} and {D. Elwell}}(1970)}]{austin70}
\bibinfo{author}{\bibnamefont{{I.G. Austin}}} \bibnamefont{and}
  \bibinfo{author}{\bibnamefont{{D. Elwell}}}, \bibinfo{journal}{Contempt.
  Phys.} \textbf{\bibinfo{volume}{11}}, \bibinfo{pages}{455}
  (\bibinfo{year}{1970}).

\bibitem[{\citenamefont{Sun et~al.}(1999)\citenamefont{Sun, Abraham, Rao, and
  Eom}}]{sun99}
\bibinfo{author}{\bibfnamefont{J.}~\bibnamefont{Sun}},
  \bibinfo{author}{\bibfnamefont{D.}~\bibnamefont{Abraham}},
  \bibinfo{author}{\bibfnamefont{R.}~\bibnamefont{Rao}}, \bibnamefont{and}
  \bibinfo{author}{\bibfnamefont{C.}~\bibnamefont{Eom}},
  \bibinfo{journal}{Appl. Phys. Lett.} \textbf{\bibinfo{volume}{74}},
  \bibinfo{pages}{3017} (\bibinfo{year}{1999}).

\bibitem[{\citenamefont{Bibes et~al.}(2001)}]{bibes2001e}
\bibinfo{author}{\bibfnamefont{M.}~\bibnamefont{Bibes}},
\bibinfo{author}{\bibfnamefont{Ll.}~\bibnamefont{Balcells}},
  \bibinfo{author}{\bibfnamefont{S.}~\bibnamefont{Valencia}},
  \bibinfo{author}{\bibfnamefont{J.}~\bibnamefont{Fontcuberta}},
  \bibinfo{author}{\bibfnamefont{M.}~\bibnamefont{Wojcik}},
  \bibinfo{author}{\bibfnamefont{E.}~\bibnamefont{Jedryka}},
  \bibnamefont{and} \bibinfo{author}{\bibfnamefont{S.}~\bibnamefont{Nadolski}},
  \bibinfo{journal}{Phys. Rev. Lett.} \textbf{\bibinfo{volume}{87}},
  \bibinfo{pages}{067210} (\bibinfo{year}{2001}).

\bibitem[{\citenamefont{Bibes et~al.}(2002)}]{bibes2002}
\bibinfo{author}{\bibfnamefont{M.}~\bibnamefont{Bibes}},
  \bibinfo{author}{\bibfnamefont{S.}~\bibnamefont{Valencia}},
  \bibinfo{author}{\bibfnamefont{Ll.}~\bibnamefont{Balcells}},
  \bibinfo{author}{\bibfnamefont{B.}~\bibnamefont{Mart\'{\i}nez}},
  \bibinfo{author}{\bibfnamefont{J.}~\bibnamefont{Fontcuberta}},
  \bibinfo{author}{\bibfnamefont{M.}~\bibnamefont{Wojcik}},
  \bibinfo{author}{\bibfnamefont{S.}~\bibnamefont{Nadolski}},
  \bibnamefont{and} \bibinfo{author}{\bibfnamefont{E.}~\bibnamefont{Jedryka}},
  \bibinfo{journal}{Phys. Rev. B} \textbf{\bibinfo{volume}{66}},
  \bibinfo{pages}{134416} (\bibinfo{year}{2002}).

\bibitem[{\citenamefont{Soeya et~al.}(2002)\citenamefont{Soeya, Hakayama,
  Takahashi, Ito, Yamamoto, Kita, Asano, and Matsui}}]{soeya2002}
\bibinfo{author}{\bibfnamefont{S.}~\bibnamefont{Soeya}},
  \bibinfo{author}{\bibfnamefont{J.}~\bibnamefont{Hakayama}},
  \bibinfo{author}{\bibfnamefont{H.}~\bibnamefont{Takahashi}},
  \bibinfo{author}{\bibfnamefont{K.}~\bibnamefont{Ito}},
  \bibinfo{author}{\bibfnamefont{C.}~\bibnamefont{Yamamoto}},
  \bibinfo{author}{\bibfnamefont{A.}~\bibnamefont{Kita}},
  \bibinfo{author}{\bibfnamefont{H.}~\bibnamefont{Asano}}, \bibnamefont{and}
  \bibinfo{author}{\bibfnamefont{M.}~\bibnamefont{Matsui}},
  \bibinfo{journal}{Appl. Phys. Lett.} \textbf{\bibinfo{volume}{80}},
  \bibinfo{pages}{823} (\bibinfo{year}{2002}).

\bibitem[{\citenamefont{Eerenstein et~al.}(2003)\citenamefont{Eerenstein,
  Kalev, Nielsen, {T.T.M. Palstra}, and Hibma}}]{eerenstein2003}
\bibinfo{author}{\bibfnamefont{W.}~\bibnamefont{Eerenstein}},
  \bibinfo{author}{\bibfnamefont{L.}~\bibnamefont{Kalev}},
  \bibinfo{author}{\bibfnamefont{L.}~\bibnamefont{Nielsen}},
  \bibinfo{author}{\bibnamefont{{T.T.M. Palstra}}}, \bibnamefont{and}
  \bibinfo{author}{\bibfnamefont{T.}~\bibnamefont{Hibma}}, \bibinfo{journal}{J.
  Magn. Magn. Mater.} \textbf{\bibinfo{volume}{258-259}}, \bibinfo{pages}{73}
  (\bibinfo{year}{2003}).

\bibitem[{\citenamefont{{J.-F. Bobo} et~al.}(2002)}]{bobo2002}
\bibinfo{author}{\bibnamefont{{J.-F. Bobo}}},
 \bibinfo{author}{\bibnamefont{{D. Basso}}},
 \bibinfo{author}{\bibnamefont{{E. Snoeck}}},
 \bibinfo{author}{\bibnamefont{{C. Gatel}}},
 \bibinfo{author}{\bibnamefont{{D. Hrabovsky}}},
 \bibinfo{author}{\bibnamefont{{J.-L. Gauffier}}},
 \bibinfo{author}{\bibnamefont{{L. Ressier}}},
 \bibinfo{author}{\bibnamefont{{R. Mamy}}},
 \bibinfo{author}{\bibnamefont{{S. Visnovsky}}},
 \bibinfo{author}{\bibnamefont{{J. Hamrle}}},
 \bibinfo{author}{\bibnamefont{{J. Teillet}}}, \bibnamefont{and}
  \bibinfo{author}{\bibnamefont{{A.R. Fert}}},
  \bibinfo{journal}{Eur. Phys. J. B} \textbf{\bibinfo{volume}{24}},
  \bibinfo{pages}{43} (\bibinfo{year}{2002}).

\bibitem[{\citenamefont{{G.S.D. Beach} et~al.}(2000)\citenamefont{{G.S.D.
  Beach}, {F.T. Parker}, {D.J. Smith}, {P.A. Crozier}, and {A.E.
  Berkowitz}}}]{beach2003}
\bibinfo{author}{\bibnamefont{{G.S.D. Beach}}},
  \bibinfo{author}{\bibnamefont{{F.T. Parker}}},
  \bibinfo{author}{\bibnamefont{{D.J. Smith}}},
  \bibinfo{author}{\bibnamefont{{P.A. Crozier}}}, \bibnamefont{and}
  \bibinfo{author}{\bibnamefont{{A.E. Berkowitz}}}, \bibinfo{journal}{Phys.
  Rev. Lett.} \textbf{\bibinfo{volume}{91}}, \bibinfo{pages}{267201}
  (\bibinfo{year}{2000}).

\bibitem[{\citenamefont{Lyonnet et~al.}(2000)\citenamefont{Lyonnet, Maurice,
  H\"ytch, Michel, and Contour}}]{lyonnet2000}
\bibinfo{author}{\bibfnamefont{R.}~\bibnamefont{Lyonnet}},
  \bibinfo{author}{\bibfnamefont{J.-L.} \bibnamefont{Maurice}},
  \bibinfo{author}{\bibfnamefont{M.~J.} \bibnamefont{H\"ytch}},
  \bibinfo{author}{\bibfnamefont{D.}~\bibnamefont{Michel}}, \bibnamefont{and}
  \bibinfo{author}{\bibfnamefont{J.-P.} \bibnamefont{Contour}},
  \bibinfo{journal}{Appl. Surf. Sci.} \textbf{\bibinfo{volume}{162-163}},
  \bibinfo{pages}{245} (\bibinfo{year}{2000}).

\bibitem[{\citenamefont{Bertacco et~al.}(2004)}]{bertacco2004}
\bibinfo{author}{\bibfnamefont{R.}~\bibnamefont{Bertacco}},
\bibinfo{journal}{unpublished}  (\bibinfo{year}{2004}).

\bibitem[{\citenamefont{{F.C. Voogt} et~al.}(1999)\citenamefont{{F.C. Voogt},
  Fujii, {P.J.M. Smulders}, Niesen, {M.A. James}, and Hibma}}]{voogt99}
\bibinfo{author}{\bibnamefont{{F.C. Voogt}}},
  \bibinfo{author}{\bibfnamefont{T.}~\bibnamefont{Fujii}},
  \bibinfo{author}{\bibnamefont{{P.J.M. Smulders}}},
  \bibinfo{author}{\bibfnamefont{L.}~\bibnamefont{Niesen}},
  \bibinfo{author}{\bibnamefont{{M.A. James}}}, \bibnamefont{and}
  \bibinfo{author}{\bibfnamefont{T.}~\bibnamefont{Hibma}},
  \bibinfo{journal}{Phys. Rev. B} \textbf{\bibinfo{volume}{60}},
  \bibinfo{pages}{11193} (\bibinfo{year}{1999}).

\bibitem[{\citenamefont{Bibes et~al.}(2004)\citenamefont{Bibes, L\"uders,
  Barth\'el\'emy, Bouzehouane, {J.-L. Maurice}, {J.-F. Bobo}, and
  Fontcuberta}}]{bibes2004}
\bibinfo{author}{\bibfnamefont{M.}~\bibnamefont{Bibes}},
  \bibinfo{author}{\bibfnamefont{U.}~\bibnamefont{L\"uders}},
  \bibinfo{author}{\bibfnamefont{A.}~\bibnamefont{Barth\'el\'emy}},
  \bibinfo{author}{\bibfnamefont{K.}~\bibnamefont{Bouzehouane}},
  \bibinfo{author}{\bibnamefont{{J.-L. Maurice}}},
  \bibinfo{author}{\bibnamefont{{J.-F. Bobo}}}, \bibnamefont{and}
  \bibinfo{author}{\bibfnamefont{J.}~\bibnamefont{Fontcuberta}},
  \bibinfo{journal}{Proc. of the 9th International Conference on Ferrites, San
  Francisco, Aug. 23-27}  (\bibinfo{year}{2004}).

\bibitem[{\citenamefont{{J. H. Scofield}}(1976)}]{scofield76}
\bibinfo{author}{\bibnamefont{{J. H. Scofield}}}, \bibinfo{journal}{J. Electron
  Spectrosc. Relat. Phenom.} \textbf{\bibinfo{volume}{8}}, \bibinfo{pages}{129}
  (\bibinfo{year}{1976}).

\bibitem[{\citenamefont{{A.N. Cormack} et~al.}(1988)\citenamefont{{A.N.
  Cormack}, {G.V. Lewis}, {S.C. Parker}, and {C.R.A. Catlow}}}]{cormack88}
\bibinfo{author}{\bibnamefont{{A.N. Cormack}}},
  \bibinfo{author}{\bibnamefont{{G.V. Lewis}}},
  \bibinfo{author}{\bibnamefont{{S.C. Parker}}}, \bibnamefont{and}
  \bibinfo{author}{\bibnamefont{{C.R.A. Catlow}}}, \bibinfo{journal}{J. Phys.
  Chem. Solids} \textbf{\bibinfo{volume}{49}}, \bibinfo{pages}{53}
  (\bibinfo{year}{1988}).

\bibitem[{\citenamefont{Tsukimura et~al.}(1997)\citenamefont{Tsukimura, Sasaki,
  and Kimizuka}}]{tsukimura97}
\bibinfo{author}{\bibfnamefont{K.}~\bibnamefont{Tsukimura}},
  \bibinfo{author}{\bibfnamefont{S.}~\bibnamefont{Sasaki}}, \bibnamefont{and}
  \bibinfo{author}{\bibfnamefont{N.}~\bibnamefont{Kimizuka}},
  \bibinfo{journal}{Jpn. J. Appl. Phys.} \textbf{\bibinfo{volume}{36}},
  \bibinfo{pages}{3609} (\bibinfo{year}{1997}).

\bibitem[{\citenamefont{{J. M. Robertson} and {A. J.
  Pointon}}(1966)}]{robertson66}
\bibinfo{author}{\bibnamefont{{J. M. Robertson}}} \bibnamefont{and}
  \bibinfo{author}{\bibnamefont{{A. J. Pointon}}}, \bibinfo{journal}{Solid
  State Comm.} \textbf{\bibinfo{volume}{4}}, \bibinfo{pages}{257}
  (\bibinfo{year}{1966}).

\bibitem[{\citenamefont{{K.J. Kim} et~al.}(2002)\citenamefont{{K.J. Kim}, {H.S.
  Lee}, {M.H. Lee}, and {S.H. Lee}}}]{kim2002}
\bibinfo{author}{\bibnamefont{{K.J. Kim}}}, \bibinfo{author}{\bibnamefont{{H.S.
  Lee}}}, \bibinfo{author}{\bibnamefont{{M.H. Lee}}}, \bibnamefont{and}
  \bibinfo{author}{\bibnamefont{{S.H. Lee}}}, \bibinfo{journal}{J. Appl. Phys.}
  \textbf{\bibinfo{volume}{91}}, \bibinfo{pages}{9974} (\bibinfo{year}{2002}).

\bibitem[{\citenamefont{{C.N. Chinnasamy} et~al.}(2001)\citenamefont{{C.N.
  Chinnasamy}, {A. Narayanasamy}, Ponpandian, Chattopadhyay, Shinoda,
  Jeyadevan, Tohji, Nakatsuka, Furubayashi, and Nakatani}}]{chinnasamy2001}
\bibinfo{author}{\bibnamefont{{C.N. Chinnasamy}}},
  \bibinfo{author}{\bibnamefont{{A. Narayanasamy}}},
  \bibinfo{author}{\bibfnamefont{N.}~\bibnamefont{Ponpandian}},
  \bibinfo{author}{\bibfnamefont{K.}~\bibnamefont{Chattopadhyay}},
  \bibinfo{author}{\bibfnamefont{K.}~\bibnamefont{Shinoda}},
  \bibinfo{author}{\bibfnamefont{B.}~\bibnamefont{Jeyadevan}},
  \bibinfo{author}{\bibfnamefont{K.}~\bibnamefont{Tohji}},
  \bibinfo{author}{\bibfnamefont{K.}~\bibnamefont{Nakatsuka}},
  \bibinfo{author}{\bibfnamefont{T.}~\bibnamefont{Furubayashi}},
  \bibnamefont{and} \bibinfo{author}{\bibfnamefont{I.}~\bibnamefont{Nakatani}},
  \bibinfo{journal}{Phys. Rev. B} \textbf{\bibinfo{volume}{63}},
  \bibinfo{pages}{184108} (\bibinfo{year}{2001}).

\bibitem[{\citenamefont{{Z.H. Zhou} et~al.}(2002)\citenamefont{{Z.H. Zhou},
  {J.M. Xue}, Wang, {H.S.O. Chan}, Yu, and {Z.X. Shen}}}]{zhou2002}
\bibinfo{author}{\bibnamefont{{Z.H. Zhou}}},
  \bibinfo{author}{\bibnamefont{{J.M. Xue}}},
  \bibinfo{author}{\bibfnamefont{J.}~\bibnamefont{Wang}},
  \bibinfo{author}{\bibnamefont{{H.S.O. Chan}}},
  \bibinfo{author}{\bibfnamefont{T.}~\bibnamefont{Yu}}, \bibnamefont{and}
  \bibinfo{author}{\bibnamefont{{Z.X. Shen}}}, \bibinfo{journal}{J. Appl.
  Phys.} \textbf{\bibinfo{volume}{91}}, \bibinfo{pages}{6015}
  (\bibinfo{year}{2002}).

\end{thebibliography}

 \end{document}